\documentstyle[aps,prl,preprint,epsf]{revtex}
\newcommand{\be}{\begin{equation}}
\newcommand{\ee}{\end{equation}}
\newcommand{\bea}{\begin{eqnarray}}
\newcommand{\beas}{\begin{eqnarray*}}
\newcommand{\eea}{\end{eqnarray}}
\newcommand{\eeas}{\end{eqnarray*}} 
\newcommand{\ba}{\begin{array}}
\newcommand{\ea}{\end{array}}
\newcommand{\bi}{\begin{itemize}}
\newcommand{\ei}{\end{itemize}}
\newcommand{\ben}{\begin{enumerate}}
\newcommand{\een}{\end{enumerate}}

\begin{document}


\title{ One Step Non SUSY Unification.}
\author{ Abdel P\'erez--Lorenzana$^{a}$
\and William A. Ponce$^{b}$ and Arnulfo Zepeda$^{a}$}
\address{
a Departamento de F\'{\i}sica,
Centro de Investigaci\'on y de Estudios Avanzados del I.P.N.\\
Apdo. Post. 14-740, 07000, M\'exico, D.F., M\'exico.\\
b Departamento de F\'\i sica, Universidad de Antioquia,  
 A.A. 1226, Medell\'\i n, Colombia.}
\date{\today}

\maketitle

\begin{abstract}
We show that it is possible to achieve one step gauge coupling
unification in a general class of non supersymmetric models which at low
energies have only the standard particle content and extra Higgs fields
doublets.
The  constraints are the experimental values of $\alpha_{em}$,
$\alpha_s$ and $\sin^2\theta_W$ at $10^2\, GeVs$,  and  the lower bounds
for FCNC and proton decay rates. Specific example are pointed out.\\[1ex]
\end{abstract}
\vskip2em

Although the Standard Model (SM) is a successful theory which is in good
agreement with experiments\cite{pdg}, it is a common belief
that there must exist a more fundamental theory, not far away from the
present experimental energies, capable to provide information to the
several aspects unanswered in the SM, in special to the so called flavor
problem which is related to the fermion mass spectrum and mixing angles,
and to the number of families in nature. The two most popular trends in
this direction in todays literature are Supersymmetry (SUSY)\cite{susy},
and Grand Unified Theories (GUTs)\cite{guts} with and without
SUSY. The hope is that the extra symmetry provides the lacking 
information. 

A well known fact nowadays is that the measured values of the SM
coupling constants at the $m_Z$ scale and the bounds on proton
life time, rule out models like 
minimal $SU(5)$\cite{georgi}, and other models that contain $SU(5)$ as an
intermediate stage in the symmetry braking chain. Another well known
result (somehow related to the analysis we are going to present next) is
that SUSY is a sufficient ingredient in order to achieve one step 
unification in GUT models\cite{amaldi}. 

In what follows we are going to show that one step unification is also
possible in a class of non SUSY GUT models. We restrict our analysis to 
models in which the low energy matter
consists only of the standard particle content and more SM Higgs doublet
fields. Our analysis excludes at the same time some of the most popular
GUT models. 

In the SM the coupling constants are defined as effective parameters which 
include  loop corrections in the gauge boson propagators according to
the renormalization group equations (rge). They are therefore energy scale
dependent, and to one loop they read
\be
\mu{d\alpha_i\over d\mu} \simeq -b_i \alpha_i^2, \label{rge}
\ee
where $\mu$ is the energy at which the coupling constants $\alpha_i =
g_i^2/4\pi$ are evaluated, with $g_1$, $g_2$, and $g_3$ the
coupling constants of the 
SM factor groups $U(1)_Y$, $SU(2)_L$ and $SU(3)_c$ respectively. The
constants $b_i$ are completely determinated by the  particle content in the
model by
\be
4\pi b_i = {11\over 3} C_i(vectors) - {2\over 3}C_i(fermions)
-{1\over 3}C_i(scalars), \label{bethas}
\ee
where $C_i(\cdots)$ the index of the representation to which the
$(\cdots)$
particles are assigned, and where we are considering Weyl fermion and
complex scalar fields. The boundary conditions at the
$m_Z\simeq 10^2 GeV$
scale for these equations are determined by the relationships
\be
\alpha^{-1}_{em} = \alpha_1^{-1}+ \alpha_2^{-1},  \quad\mbox{and}\quad
\tan^2\theta_W = {\alpha_1\over\alpha_2}, \label{rel1}
\ee
valid at all energy  scales, and by the experimental values 
\bea
 \alpha^{-1}_{em} &=& 127.90 \pm 0.09~\cite{pdg,aem},\nonumber\\  
\sin^2\theta_W &=& 0.2315 \pm 0.0002~\cite{pdg} \quad \mbox{and}\\
\alpha_3 &=& \alpha_s = 0.1123\pm 0.006~\cite{pdg,as}.\nonumber
\eea

The unification of the SM gauge coupling constants is achieved
if they merge together into a common value $\alpha=g^2/4\pi$ at a
certain energy scale $M$, where $g$ is the gauge coupling constant of
the unifying group $G$. However, since $G\supset G_s$,  
the normalization of the  generators corresponding  to the subgroups
$U(1)_Y$, $SU(2)_L$ and $SU(3)_c$ is in general different for
each particular group $G$, and therefore the SM  coupling constants
$\alpha_i$ differ at the  
unification scale from $\alpha$ by numerical factors $c_i\,(\alpha_i
=c_i\alpha )$ which are pure rational numbers 
satisfying $c_i\leq 1$ (due to the normalization of the generators in
$G$). For example in $SU(5)$, $c_1 = {3\over 5}$ and $c_2 = c_3 =1$. These
values are the same in
$SO(10)$~\cite{so10} and $E_6$~\cite{e6}, but they are different for 
other cases which do not contain $G_s$ embedded into an $SU(5)$
subgroup~\cite{guts} as it is the case for $E_7$~\cite{e7},
$SU(5)\otimes SU(5)$~\cite{2su5}, $[SU(6)]^3\times Z_3$~\cite{3su6},
$[SU(6)]^4\times Z_4$~\cite{4su6}, $SU(8)\otimes SU(8)$~\cite{2su8} or
the Pati-Salam models~\cite{patis}.

The constants $c_i$ are fixed once we fix the unifying gauge structure.
Then, from eq.(\ref{rel1}) it follows that at the unification scale the
value
of $\sin^2\theta_W$ is given by
\be
\sin^2\theta_W = {\alpha_{em}\over\alpha_2} = {c_1\over c_1 + c_2}.
\ee

In this paper we shall consider
for $c_3$ only two values, $c_3=1$ for those models which contain
$SU(3)_c$
embedded into a simple group, or $c_3={1\over 2}$ for those which contain  
$SU(3)_c$ embedded into the chiral color extension $SU(3)_{cL}\otimes
SU(3)_{cR}$~\cite{2su3}.

To compute the $b_i$ coefficients in the rge we will assume
that only the standard particles are light so that, according to the  
decoupling  theorem~\cite{appel}, only they contribute. We obtain 
\be 
2 \pi \left( \ba{c} b_1\\[1ex] b_2\\[1ex] b_3 \ea \right) = 
 \left( \ba{c} 0\\[1ex] 22\over 3 \\[1ex] 11 \ea \right) - 
 \left( \ba{c} 20\over 9 \\[1ex] 4\over 3 \\[1ex] 4 \over 3 \ea \right) F - 
 \left( \ba{c}  1\over 6 \\[1ex]   1\over 6 \\[1ex] 0 \ea \right) H,
\ee
where $F$ is the number of families and $H$ is the number of low energy 
complex
Higgs doublets (whose contribution was neglected in the early analysis, 
see for example the first references in  ~\cite{georgi,so10,e6,2su5}). 
Notice that we are not including in the former equation the
normalization factor $3\over 5$ into $b_1$ coming from
the $SU(5)$ theory and wrongly included in some general discussions. In the
minimal SM, $F=3$ and $H=1$. Nevertheless, a more general model could have  
more than one low energy Higgs field doublet, then $H$ may be taken as a
free parameter. Notice also that we are including in our
analysis only doublet Higgs fields, due to the facts that singlets do
not
contribute to the rge, and the presence of higher multiplets may spoil
the $\Delta I=1/2$ weak isospin rule.

The solutions to (\ref{rge}) are
\be
\alpha_i^{-1}(m_Z) = {1\over c_i}\alpha^{-1}- b_i(F,H)\, \ln\left({M\over
m_Z}\right), \label{system}
\ee
which for $i=1,2,3$ constitute a system of three equations with the
unification variables $\alpha$, $M$ and $H$ as the three unknowns
(for $F=3$ families). The system of Eqs. (\ref{system}) may be 
solved for these variables in function
of the numerical factors $c_i$ and the experimental values for
$\alpha_i$ at the $m_Z$ scale. The solution is unique for each set
 of values $\{c_1,c_2,c_3\}$ characteristic of each model.
For $c_3 = 1$ (and also for $c_3=\frac{1}{2}$), the solutions to 
(\ref{system}) produce 
three families of curves in the $c_1-c_2$ plane defined by the equations
$\alpha(c_1,c_2) = c_\alpha$, $H(c_1,c_2) = c_H$ and $M(c_1,c_2) = c_M$,
where $c_\alpha, c_H$ and $c_M$ are arbitrary constant values. Each 
curve in each family is then characterized by the numerical constant
$c_\alpha, c_H$ and $c_M$ respectively. As a
consequence, for each point in the plane $(c_1,c_2)$ corresponds unique
values for the unification variables associated to those curves which
intersect at that particular point. 

Now, there exist some experimental and theoretical bounds for the
possible values of the unification variables. 
First, the unification scale $M$ must be lower than the Plank scale
$M_{P}\sim G_N^{1/2}\sim 10^{19}GeV$, and also it must
be greater than $10^5GeV$ in order to agree with the experimental
bounds on FCNC~\cite{pdg}. Also, 
since some models predict proton decay, and the experimental bound for
the proton life time $\tau_p$ is  
$\tau_{p\rightarrow e\pi}\sim M^4 >10^{32}$ Yrs, then 
$M$ must be greater than $10^{16} GeV$ if the proton is unstable in the
model under consideration. Hence, in the analysis we have to consider two
different zones in the $c_1-c_2$ plane, given by $10^{16}GeV<M<M_{P}$
and
$10^5GeV \leq M \leq 10^{16}GeV$, which admit and does not admit proton
decay respectively.  Next, because $b_3>0$ and $b_1<0$ always,
$\alpha_1(m_Z)<\alpha<\alpha_s(m_Z)/c_3$ and thus $\alpha$ is finite. 
Hence, as $\ln (M/m_Z)$ is also finite, from (\ref{system}) we deduce that
$H$ should be also finite and then there is an upper bound $H_{max}$
which represents the maximum number of low energy Higgs doublets allowed. 
Therefore, $0\leq H\leq H_{max}$. 
These bounds limit the region in the $c_1-c_2$ plane where the coupling
constant unification  is possible and consistent  with the experimental
data and theoretical requirements. Notice also that $H$ can take only 
integer values.

The solutions of eqs. (\ref{system}) for $\alpha, H$ and $M$ are:  
\be
\alpha^{-1} =c_1c_2c_3\cdot 
{(\alpha_1^{-1}-\alpha_2^{-1})(99 - 12F) + \alpha_3^{-1}(8F + 66)
\over c_1c_2(8F + 66) + c_1 (c_1-c_2)(12F - 99)},
\ee
\be
H = {2\over 3}\cdot {c_2(\alpha_1^{-1}c_1 - \alpha_3^{-1}c_3)(66-12F) + 
c_3(\alpha_1^{-1} c_1 - \alpha_2^{-1}c_2)(12F-99) +
20c_1(\alpha_2^{-1}c_2 - \alpha_3^{-1}c_3)
\over c_1c_2(\alpha_1^{-1}-\alpha_2^{-1}) + \alpha_3^{-1}c_3 (c_1-c_2)},
\ee
\be
\ln\left({M\over m_Z}\right) = 
18\pi\cdot {c_1c_2(\alpha_1^{-1}-\alpha_2^{-1}) + \alpha_3^{-1}
c_3 (c_1-c_2)\over c_1c_2(8F + 66) + c_1 (c_1-c_2)(12F - 99)}.
\ee
From these expressions, the limited region obtained for values of $c_1$ and
$c_2$ that give unification is plotted in figure 1 for $c_3=1$ and in 
figure 2 for $c_3={1\over2}$, where we used $F=3$ for three families,
and central values for  $\alpha_s$, $\alpha_{em}$ and
$\sin^2\theta_W$. Let us see the consequences of those graphs:

\noindent
{\bf Analysis of Fig. 1}: It corresponds to the case of a GUT group which
does not include chiral color symmetries. The allowed region of parameters
$(c_1,c_2)$ lies inside the lines $M=10^5$ GeVs, $H=0$ and $c_2=1$.
There is a maximum unification mass scale possible given by 
$M\leq 10^{17.5}$ GeVs $< M_P$ and the number of Higgs field doublets
allowed is such that $0<H\leq 91$ in general, but if the proton does decay
in the context of the GUT model then $0<H\leq 2$.
Let us see the implications of this for some specific models:\\
{\bf 1-$SU(5)$}. For all the models in this group proton decay
is always present~\cite{georgi}, and $(c_1,c_2)=({3\over 5},1)$ which lies
inside the allowed zone,
but in a region where $M=10^{13}$GeVs. Hence the $SU(5)$ GUT scale $M$ is
in conflict with the bounds for proton decay. Since $SU(5)$ allows only
one step symmetry breaking chain (sbc) 
$SU(5)\stackrel{M}{\longrightarrow}SM$, $SU(5)$ is ruled out in
general. That is, the experimental bounds on proton decay rule out not
only minimal $SU(5)$ but also all the possible extensions which include
arbitrary representations of Higgs field multiplets.\\
{\bf 2-$SO(10)$}. Like for the previous model, proton decay is always
present for this group~\cite{so10} and $(c_1,c_2)=({3\over 5},1)$.
Therefore the
one step sbc $SO(10)\stackrel{M}{\longrightarrow}SM$ is ruled out. From
our analysis nothing can be say about the two stage sbc 
$SO(10)\stackrel{M}{\longrightarrow}SU(4)_c\otimes SU(2)_L\otimes SU(2)_R
\stackrel{M^\prime}{\longrightarrow}SM$.\\
{\bf 3-$SU(4)_c\otimes SU(2)_L\otimes SU(2)_R$}. This group can be viewed 
as a subgroup of $SO(10)$~\cite{so10}, or as a subgroup of the 
Pati-Salam model~\cite{patis}, or either as a non-simple unification   
model by
itself. For this group $(c_1,c_2)=({3\over 5},1)$ again. In this model,
proton can not decay via leptoquark gauge bosons (see the first paper in
~\cite{patis}), but it can decay via Higgs fields scalars. So, the one
stage breaking of this model is not ruled out as long as one can break the
symmetry using scalars which do not break spontaneously the baryon
quantum number B.\\
{\bf 4-$E_6$}. Proton decay is always present for this 
group~\cite{e6}, and $(c_1,c_2)=({3\over 5},1)$ also. So, the one step sbc
$E_6\stackrel{M}{\longrightarrow}SM$ is ruled out. Nothing can be said for
the multistage sbc.\\
{\bf 5-$SU(3)_L\otimes SU(3)_c\otimes SU(3)_R$}. This group can be viewed  
as a subgroup of $E_6$~\cite{e6} or as a unification model by itself (the
trinification model of Georgi-Glashow-de Rujula~\cite{rug}). Again
$(c_1,c_2)=({3\over 5},1)$ and the proton decay in the model is only
Higgs-boson mediated. {\it The one stage breaking of
this model is not ruled out} as long as one can break the symmetry using
scalars which do not break spontaneously B (see the second paper in
~\cite{rug}).\\
{\bf 6-$SO(18)$}. Proton decay is always present for this group, and
$(c_1,c_2)=({3\over 5},1)$ ~\cite{so18}. The conclussions are the same than
for
$E_6$.\\
{\bf 7-$[SU(6)]^3\times Z_3$}. The proton is stable in the
context of this model\cite{3su6}. For this group 
$(c_1,c_2)=({3\over 14}, {1\over 3})$ which lies
outside the allowed zone and one stage sbc is ruled out (the two stage
sbc for the model is presented in the last paper of Ref.~\cite{3su6}).

\noindent
{\bf Analysis of Fig. 2}: It corresponds to the case of a GUT group which
includes chiral color symmetries. The allowed region in the plane 
$(c_1,c_2)$ lies inside the lines $M=10^5$ GeVs, $H=0$, $M=10^{19}$ GeVs
=$M_P$ and $c_2=1$. Therefore there is no bound for a maximum unification
mass scale
and the allowed number of Higgs field doublets is such that $0<H\leq 136$
in general, but if the proton does decay in the context of the GUT model
then $0<H\leq 28$. Let us see the implications of this for some specific
models:\\
{\bf 1-$SU(5)\otimes SU(5)$}. Proton decay is mediated via gauge and
Higgs bosons for the models in this group~\cite{2su5}. 
$(c_1,c_2)=({3\over 13},1)$
which lies inside the allowed zone but in a region where $M<<10^{16}$
GeVs, in serious conflict with bounds for proton decay. The models
are all ruled out.\\
{\bf 2-$[SU(6)]^4\times Z_4$}. The proton is stable in the context of
the model presented in Ref. ~\cite{4su6}. For this group 
$(c_1,c_2)=({3\over 19} , {1\over 3})$ which lies inside the allowed zone.
So,
{\it one stage sbc for this model is also possible}, and it is presented
in Ref.~\cite{4su6}.

We mention that our analysis has been done 
assuming non supersymmetric unification. Also we have neglected
thresholds effects which depend on the particular structure of each
model, we do not include second order corrections to the rge which are
typically of the order of 1 to 10\%, and we have not included the
experimental errors of the SM gauge coupling constants. 

The previous analysis allows us to conclude that it is indeed possible to
achieve the unification of the coupling constants of the SM in one step in
a general class of non supersymmetric models. Two particular models
with simple unifying groups were
single out: the trinification model of Georgy-Glashow-de Rujula\cite{rug}
for GUT groups which do not include chiral color symmetry, and the model
in Ref.~\cite{4su6} for GUT models with chiral color symmetry. 

\noindent
{\bf ACKNOWLEDGEMENTS}
The authors acknowledge P. Langacker and G. Kane for discussions and 
comments. This work was partially supported by CONACyT, M\'exico and
COLCIENCIAS, Colombia.

\figure{
\centerline{
\epsfxsize=260pt
\epsfbox{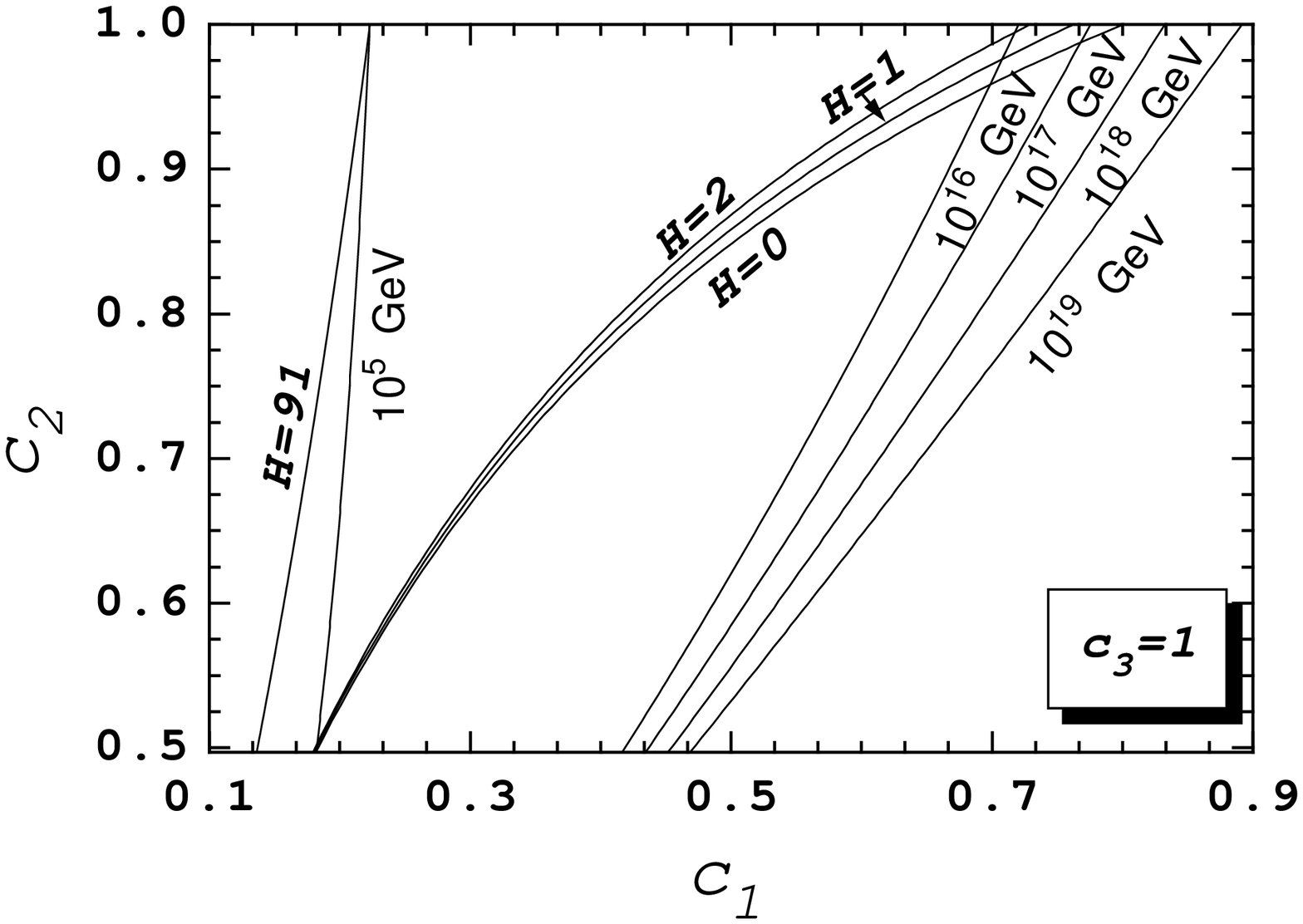}
}\caption{Plots for some values of H and M for the non chiral
color models. The bounds in $c_1$ and $c_2$, impose at once for $\alpha$
the bounds $16.5454< \alpha^{-1} < 48.4809$.}
}
\vskip1em
\figure{
\centerline{
\epsfxsize=260pt
\epsfbox{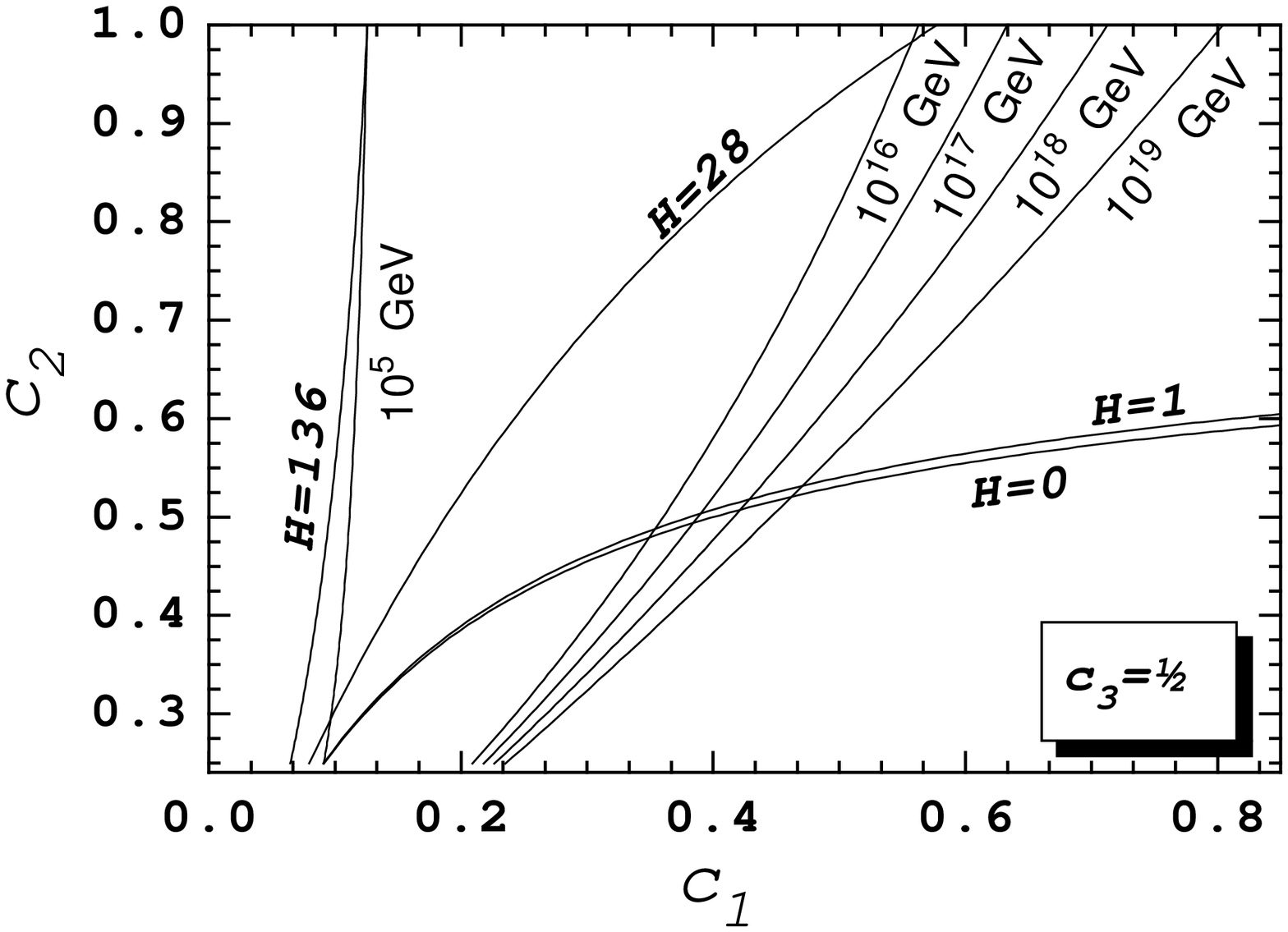}
}\caption{Plots for some values of $H$ and $M$
for GUTs containing the chiral color extension. In this case we have
$8.27269< \alpha^{-1}< 26.1967$.}
}


\begin{thebibliography}{99}
\bibitem{pdg} 
Particle  Data Group:
 L. Montanet {\it et al.}, \prd{\bf 50}, 1173 (1994).
 R.M. Barnett {\it et al.}, \prd{\bf 54}, 1 (1996).
\bibitem{susy}
For a collection of reports on SUSY see: ``{\it Supersymmetry and
Supergravity, A reprint volume of Physics Reports}". Edited by M. Jacob
(North Holland/World Scientific, Amsterdam/Singapore 1986).
\bibitem{guts} 
For a Review see 
G. G. Ross {\it Grand Unified Theories.} Frontiers in physics; No. 60
(Benjamin-Cumings Pub. Co., 1984).
R. Mohapatra, {\it Unification and Supersymmetry.} 2nd. ed. (Springer--Verlag,
Berlin, 1992). 
P. Langacker, {\it Phys. Rep.} {\bf 72}, 185 (1981), and references therein.
\bibitem{georgi} 
H. Georgi and S. L. Glashow, \prl{\bf 32}, 438 (1974). 
H. Georgi, H. R. Quinn and S. Weimberg \prl{\bf 33}, 451 (1974).
\bibitem{amaldi}
U.Amaldi, W.de Boer and H. Furstenau, Phys. Lett {\bf B260}, 447 (1991);
P.Langacker and M. X. Luo, Phys. Rev. {\bf D44}, 817 (1991); C.Giunti,
C.W. Kim and U.W. Lee, Mod. Phys. Lett {\bf A6}, 1745 (1991).  
\bibitem{aem} 
S. Fanchiotti, B. Kniehl, and A. Sirlin, \prd{\bf 48}, 307 (1993) and
references therein.
\bibitem{as} 
M. Virchaux and A. Milsztajn, \pl {\bf 274B}, 221 (1992).
P. Abreu {\it el al.}, (DELPHI Collab.), Z. Phys {\bf C54}, 55 (1992). 
M. Shifman, Int. J. of Mod. Phys. {\bf A11}, 3195 (1996). 
\bibitem{so10} 
H. Georgi, in {\it Particles and Fields -- 1974}. Edited by C. E. Carlson.
(American Institute of physics, New York, 1975), p. 575. H. Frizsch and P.
Minkowsky, Ann. Phys. {\bf 93}, 193 (1975). 
\bibitem{e6} 
F. G\"ursey, P. Ramond and P. Sikivie, \pl{\bf 60B}, 177 (1975).
S. Okubo, \prd{\bf 16}, 3528 (1977).
\bibitem{e7} 
F. G\"ursey and P. Sikivie, \prl{\bf 36}, 775 (1976); \prd{\bf 16}, 816
(1977). P. Ramond, Nucl. Phys. B{\bf 110}, 214 (1976); B{\bf 126}, 509 (1977).
\bibitem{2su5} 
A. Davidson and K. C. Wali, \prl{\bf 58}, 2623 (1987). 
R. N. Mohapatra, hep-ph/9601203.
\bibitem{3su6} 
W. A. Ponce, in {\it Proceedings of the Third Mexican School of Particles and
Fields.} Edited by J. L. Lucio and A. Zepeda (World  Scientific, Singapore,
1989), pp. 90-129. A. H. Galeana, R. Mart\'{\i}nez, W. A. Ponce and A. Zepeda,
\prd{\bf  44}, 2166 (1991). W. A. Ponce and A. Zepeda, \prd{\bf 48}, 240
(1993);J. B. Fl\'orez, W. A. Ponce and A. Zepeda, Phys. Rev. {\bf D49},
4958 (1994). 
\bibitem{4su6} 
W. A. Ponce and A. Zepeda, Z. Physik {\bf C63}, 339 (1994); 
A. P\'erez-Lorenzana, W. A. Ponce and A. Zepeda, {\it Nonsupersymmetric
gauge coupling unification in} $[SU(6)]^4\times Z_4$ {\it and proton
stability}, submitted for publication. 
\bibitem{2su8} 
Y. F. Pirogov, IHEP 79-72. J. C. Pati and A. Salam,  Nucl. Phys.
B{\bf 150}, 76 (1970).
\bibitem{patis} 
J. C. Pati and A. Salam, \prl{\bf 31}, 661 (1973); {\bf 36}, 1229
(1976); \pl{\bf 58B}, 333 (1975); \prd{\bf 8}, 1240 (1973); 
D{\bf 10}, 275 (1974).
\bibitem{2su3} 
P. H. Frampton and S. L. Glashow, \pl{\bf 190B}, 157 (1987);
\prl{\bf 58}, 2168 (1987).
\bibitem{appel}
T. Appelquist and J. Carazzone, \prd{\bf 11}, 2856 (1975).
\bibitem{rug}
S.L.Glashow, in {\it Proceedings of the fifth workshop on grand
Unification}. Edited by K.Kang,
H.Fried and P. Frampton (World Scientific P.C., Singapore 1994), 
pp 88-94;
Babu K., X.G.He and S. Pakvasa, Phys. Rev. {\bf D33}, 763 (1986).
\bibitem{so18}
F.Wilczek and A.Zee, Phys. Rev. {\bf D25}, 553 (1982); J. Bagger and S.
Dimopoulos, Nucl. Phys. {\bf B244}, 247 (1984); J. Bagger {\it et al.}, Nucl.
Phys. {\bf B258}, 566 (1985).

\end{thebibliography}
\end{document}